\documentclass[useAMS,usenatbib]{mn2e}
\RequirePackage{lineno}
\usepackage{graphicx}  
\usepackage{ amssymb }
\usepackage{journals}

\def\simgt{\lower.5ex\hbox{$\; \buildrel > \over \sim \;$}}
\def\simlt{\lower.5ex\hbox{$\; \buildrel < \over \sim \;$}}

\def\Teff{$T_{\rm eff}$}

\def\alfa2d{MLT--$\,\alpha^{\rm 2D}$}

\def\Dnu{$\Delta\nu$}
\def\numax{$\nu_{\rm max}$}

\def\kepler{\mbox{\textit{Kepler}}}

\title[]{Galactic archaeology: mapping and dating stellar populations with asteroseismology of red-giant stars}

\author[A. Miglio et al.]{A. Miglio$^{1,2}$\thanks{E-mail: miglioa@bison.ph.bham.ac.uk}, 
 C. Chiappini$^{3}$, T. Morel$^{4}$, M. Barbieri$^{5}$, W. J. Chaplin$^{1}$, L. Girardi$^{6}$,\newauthor J. Montalb{\'a}n$^{4}$, M. Valentini$^{4}$, B. Mosser$^{7}$, F. Baudin$^{8}$, L. Casagrande$^{9}$,\newauthor L. Fossati$^{10}$,  V. Silva Aguirre$^{11}$, A. Baglin$^{7}$\\
 \\
$^{1}$School of Physics and Astronomy, University of Birmingham, Edgbaston, Birmingham B15 2TT, United Kingdom\\
$^{2}$Kavli Institute for Theoretical Physics, Kohn Hall, University of California, Santa Barbara, CA 93106, USA\\
$^{3}$Leibniz-Institut fur Astrophysik Potsdam (AIP), An der Sternwarte 16, D-14482, Potsdam, Germany\\
$^{4}$Institut d'Astrophysique et de G\'eophysique de l'Universit\'e de Li\`ege, All\'ee du 6 Ao\^ut, 17 B-4000 Li\`ege, Belgium\\
$^{5}$Dipartimento di Astronomia, Universitˆ di Padova, Vicolo Osservatorio, 3, 35122 Padova, Italy\\
$^{6}$INAF-Osservatorio Astronomico di Padova, Vicolo dell'Osservatorio 5, I-35122 Padova, Italy\\
$^{7}$LESIA, CNRS, Universit{\'e} Pierre et Marie Curie, Universit{\'e} Denis Diderot, Observatoire de Paris, 92195 Meudon cedex, France\\
$^{8}$Institut d'Astrophysique Spatiale, Univ Paris-Sud, UMR8617, CNRS, B‰timent 121, 91405 Orsay Cedex, France\\
$^{9}$Research School of Astronomy \& Astrophysics, Mount Stromlo Observatory, The Australian National University, ACT 2611, Australia\\
$^{10}$Argelander-Institut f\"ur Astronomie der Universit\"at Bonn, Auf dem H\"ugel 71, 53121 Bonn, Germany\\
$^{11}$Stellar Astrophysics Centre, Department of Physics and Astronomy, Aarhus University, Ny Munkegade 120, DK-8000 Aarhus C, Denmark\\
}         
\begin{document}
\date{}
\pagerange{\pageref{firstpage}--\pageref{lastpage}} \pubyear{2010}

\maketitle

\label{firstpage}

\begin{abstract}
Our understanding of how the Galaxy was formed and evolves is severely hampered by the lack of precise constraints on basic stellar properties such as distances, masses, and ages. Here, we show that solar-like pulsating red giants represent a well-populated class of accurate distance indicators, spanning a large age range, which can be used to map and date the Galactic disc in the regions probed by observations made by the CoRoT\footnote{The CoRoT space mission, launched on December 27th 2006, has been developed and is operated by CNES, with the contribution of Austria, Belgium, Brazil, ESA (RSSD and Science Programme), Germany and Spain.} and \kepler\ space telescopes. When combined with photometric constraints, the pulsation spectra of such evolved stars not only reveal their radii, and hence distances, but also provide well-constrained estimates of their masses, which are reliable proxies for the ages of the stars. As a first application we consider red giants observed by CoRoT in two different parts of the Milky Way, and determine precise distances for $\sim$2000 stars spread across nearly 15,000 pc of the Galactic disc, exploring regions which are a long way from the solar neighbourhood. We find significant differences in the mass distributions of these two samples which, by comparison with predictions of synthetic models of the Milky Way, we interpret as mainly due to the vertical gradient in the distribution of stellar masses (hence ages) in the disc.  In the future, the availability of spectroscopic constraints for this sample of stars will not only improve the age determination, but also provide crucial constraints on age-velocity and age-metallicity relations at different Galactocentric radii and heights from the plane. 

\end{abstract}
\begin{keywords}
asteroseismology;  stars: distances; Galaxy: disc
\end{keywords}

\section{Introduction}
\label{sec:intro}
Accurate and precise distances to stars are a fundamental cornerstone of astronomy. Such data play a key role in shaping theories of the history and fate of our Galaxy. Another fundamental cornerstone is provided by stellar ages. Ages can be estimated only for a limited sample of individual field stars, and then only by using either model-dependent techniques or empirical methods that still need careful calibration \citep{Soderblom2010}. 
These two cornerstones are essential to Galactic Archaeology, the study of the chemical compositions and dynamic motions of stars of different ages, within which is coded information on the origin and subsequent evolution of the Milky Way \citep{Turon2008,  Freeman2002}. There is a modern consensus among astronomers that a variety of processes play a role in shaping our Galaxy, for example gas accretion and in-situ star formation, mergers, and dynamical processes in the Galaxy (such as diffusion and migration of stars). Galactic Archaeology research has now reached a point where so-called Òchemo-dynamicalÓ models are needed to quantify the relative importance of each of these processes, over the course of more than 10 Gyr of evolution of  the Milky-Way. Stringent observational constraints are in turn desperately needed to test these model predictions. 

Particularly important observational constraints are the age-velocity and age-metallicity relations at different positions in the Galaxy \citep[see e.g.][]{Freeman2002,Minchev2012}. These relations are uncertain, or limited to neighbouring stars (e.g. from the Geneva-Copenhagen Survey \citep{Holmberg2009, Casagrande2011}, which is confined to distances less than about 100 pc from us).  Future planned surveys hope to fill this gap by combining precise proper motions from the ESA-GAIA \citep{Perryman2001} astrometric mission with massive spectroscopic follow-up from ground-based telescopes, to extend the region for which distances and chemical compositions are known out to as far as 10,000 pc. However, this will not happen before the beginning of the next decade, and it will still need to confront the large and systematic uncertainties affecting the age determinations of single stars. 

Red giants are cool highly luminous stars, which, by virtue of covering a wide domain in mass, age, chemical composition, and evolutionary state, are in principle an important source of data for testing chemo-dynamical models. Previously, age estimation of red giants has had to rely on constraints on their surface properties only. A severe limitation of such an approach is that during the red-giant phase stars of significantly different age and distance end up sharing very similar observed surface properties making it extremely hard to discriminate evolutionary states of field giants belonging to the Galactic-disc population. That situation has now changed thanks to asteroseismology, with the detection of solar-like pulsations in thousands of red giants \citep{DeRidder09, Bedding10, Hekker2011b} observed by the CoRoT \citep{Baglin06a} and \kepler\ \citep{Gilliland10} space telescopes. The pulsation frequencies may be used to place tight constraints on the fundamental stellar properties, including radius, mass and evolutionary state \citep{Stello2008, Kallinger2010, Mosser2010, Montalban2010, Bedding2011},  the properties of Helium ionisation regions \citep{Miglio2010}, and internal rotation \citep[see e.g.][]{Beck2012}. Here, we show it is possible to use pulsating red giants as accurate distance indicators, and tracers of stellar population ages.

\section{Method}
\label{sec:dir}
Radii and masses of solar-like oscillating stars can be estimated from the average seismic parameters that characterise their oscillation spectra: the so-called average large frequency separation (\Dnu), and the frequency corresponding to the maximum observed oscillation power (\numax).

The large frequency spacing is predicted by theory to scale as the square root of the mean density of the star \citep[see e.g.][]{Vandakurov67, Tassoul80}:
\begin{equation}
\Delta\nu\simeq\sqrt{\frac{M/M_\odot}{(R/R_\odot)^3}}\Delta\nu_{\odot}{\rm,}
\label{eq:dnu}
\end{equation}
where $\Delta\nu_\odot=135$ $\mu$Hz. The frequency of maximum power is expected to be proportional to the acoustic cutoff frequency \citep{Brown1991, Kjeldsen1995, Mosser2010, Belkacem2011}, and therefore:
\begin{equation}
\nu_{\rm max}\simeq\frac{M/M_\odot}{(R/R_\odot)^2\sqrt{T_{\rm eff}/T_{\rm eff,\odot}}}\nu_{\rm max,\odot}\,{\rm,}
\label{eq:numax}
\end{equation}
where $\nu_{\rm max,\odot}=3100$ $\mu$Hz and $T_{\rm eff,\odot}=5777$ K.

When no information on distance/luminosity is available, which is the case for the vast majority of field stars observed by CoRoT and \kepler, Eq. \ref{eq:dnu} and \ref{eq:numax} may be solved to derive $M$ and $R$ \citep[see e.g.][]{Stello2008,Kallinger2010, Mosser2010}:
\begin{eqnarray}
\frac{M}{M_\odot} &\simeq& \left(\frac{\nu_{\rm max}}{\nu_{\rm max, \odot}}\right)^{3}\left(\frac{\Delta\nu}{\Delta\nu_{\odot}}\right)^{-4}\left(\frac{T_{\rm eff}}{T_{\rm eff, \odot}}\right)^{3/2} \label{eq:scalM}      \\
\frac{R}{R_\odot} &\simeq& \left(\frac{\nu_{\rm max}}{\nu_{\rm max, \odot}}\right)\left(\frac{\Delta\nu}{\Delta\nu_{\odot}}\right)^{-2}\left(\frac{T_{\rm eff}}{T_{\rm eff, \odot}}\right)^{1/2}{\rm.}\label{eq:scalR}
\end{eqnarray}

We determined stellar radii and masses by combining the available seismic parameters $\nu_{\rm max}$ and $\Delta\nu$ with effective temperatures $T_{\rm eff}$. The latter were determined using 2MASS \citep{Skrutskie2006} $J$ and $K_s$ photometry and the colour-$T_{\rm eff}$ calibrations by \cite{Alonso1999}, which depend only weakly on metallicity.  2MASS colours were transformed into the CIT photometric system used by \citet{Alonso1999} using the relations available in \citet{Alonso1998} and \citet{Carpenter2001}.

We then computed luminosities  $L$ using the Stefan-Boltzmann law $L=4\pi  R^2 \sigma T_{\rm eff}^4$, where $\sigma$ is the Stefan-Boltzmann constant.  The distance modulus of each star was determined as $K'_{s\rm0}-K_{s\rm0}$, where $K_{\rm s0}$ is the de-reddened apparent 2MASS $K_s$ magnitude, and $K'_{s\rm0}$ the absolute $K_s$ magnitude. The latter was obtained by combining $L$ and the bolometric corrections from \cite{Girardi05}, which are based on the Castelli \& Kurucz (2005) ATLAS9 model atmospheres, for the range of $T_{\rm eff}$ and $\log{g}$ under consideration.

We took into account the effect of interstellar extinction on the magnitude and colour of each star using the 3D model of Galactic extinction in the $V$ band ($A_{\rm V}$) by \citet{Drimmel2003}.  The extinctions in the $J$ and $K_{\rm s}$ bandpasses were determined following \citet{Fiorucci2003} assuming the spectral energy distribution of a K1 giant.
Since the extinction is distance dependent, we iterated the procedure until the derived distance does not vary by more than $1\%$.   We chose to consider magnitudes in the near-IR  to reduce the effect of interstellar reddening in both the determination of $T_{\rm eff}$ and apparent de-reddened magnitudes.

By analogy with the well-known period-luminosity relation used to estimate distances of classical pulsators \citep[see e.g.][for a review]{Feast1987, Madore1991, Bono2010}, we can explicitly write a relation between the distance and the pulsation properties of solar-like oscillators. In contrast to the case of single-mode, radial pulsators we can estimate stellar radii without making any assumptions\footnote{Excluding the weak dependence of bolometric correction on metallicity.} on mass-metallicity-luminosity relations. Hence we find:
\begin{eqnarray*}
\log{d}=1+2.5\log{\frac{T_{\rm eff}}{T_{\rm eff,\odot}}}+\log{\frac{\nu_{\rm max}}{\nu_{\rm max,\odot}}}+\\-2\log{\frac{\Delta\nu}{\Delta\nu_\odot}}+0.2(m_{\rm bol}-M_{\rm bol, \odot}) {\rm ,}
\end{eqnarray*}
where $d$ is expressed in parsec, $m_{\rm bol}$ is the apparent bolometric magnitude of the star, and $M_{\rm bol, \odot}$ the absolute solar bolometric magnitude. Although this expression is model-independent, it builds upon Eq \ref{eq:dnu} and Eq. \ref{eq:numax} which are based on simplifying assumptions that need to be independently verified, and supported with empirical tests.

Tests against independent estimates of mass and radius suggest that $R$ and $M$ derived using average asteroseismic parameters are accurate to within 5\% and 10\%, respectively (see \citealt{Miglio2013} for a review on the tests performed so far, see \citet{Miglio2012c, SilvaAguirre2012, Huber2012} for tests on nearby stars).

\section{Data and uncertainties}
The CoRoT photometric time series used in this work were obtained in the
so-called exofield during the first long CoRoT runs:  LRc01 ($l=37^\circ$, $b=-7^\circ$) and LRa01 ($l=217^\circ$, $b=-2^\circ$). These
long runs lasted approximately 140 days, providing us with a frequency
resolution of about 0.08 $\mu$Hz. A first analysis of these data was made
by \citet{DeRidder09}, \citet{Hekker09} and \citet{Kallinger2010}.
Solar-like oscillations were detected in  435 and 1626 giants belonging to LRa01 and LRc01, respectively. 
To derive $M$,$R$, and distance we use  the values (and uncertainties) of \numax\ and \Dnu\ as determined by \citet{Mosser2010}. The typical uncertainty on \numax\ and \Dnu\ for the 150-d long CoRoT observations is of the order of  2.4\% and 0.6\%, respectively.   

The uncertainties on the stellar properties are estimated using Monte Carlo simulations adopting the following constraints:
\begin{itemize}
\item Apparent magnitudes: uncertainties on $J$ and $K_s$ were taken from the 2MASS photometry available in the EXODAT \citep{Deleuil2009} catalogue (the median uncertainty is 0.02 mag). 
\item Extinction/reddening: a random error in $A_V$ of 0.3 mag was considered both in determining \Teff\ photometrically and in de-reddening $K_s$ apparent magnitudes.
\item \Teff: for each star we considered two sources of uncertainty. The first (100 K) was due to uncertainties on the colour-\Teff\ calibration itself   \citep{Alonso1999}. The second one was due to uncertainties on reddening. This resulted in a median combined uncertainty on \Teff\ of $\sim 190$ K (calibration+reddening).
\end{itemize}

When determining the distance, reddening affects not only  the de-reddened apparent  bolometric magnitude, but also the determination of $L$ (mostly via \Teff). A higher $A_V$ increases the estimated \Teff, hence $L$, but it also increases the apparent de-reddened luminosity $l$. Since $d \propto (L/l)^{1/2}$, the overall effect of reddening on the distance itself is  partly reduced.
%
 By performing 500 realisations of the data assuming the uncertainties described above to be Gaussian,  we find a median intrinsic uncertainty of 5\% in distance, 3.6 \% in radius, and 10 \% in mass. 
To explore the effect of possible systematic offsets in the \Teff\ scale \citep[see e.g.][]{Casagrande2010}, we increased/decreased  \Teff\  by 100 K, and found that the distance estimate was affected by $~2.5\%$ (the radius by 1\% and the mass by 3\%).
This leads to an overall  uncertainty on the distance of  5\% (random) + 5\% (systematic due to radius determination via scaling relations) + 2.5\% (systematic due to the \Teff\ scale).

\begin{figure*}
\vspace*{-1.7cm}

\centering
      \includegraphics[width=.37\hsize]{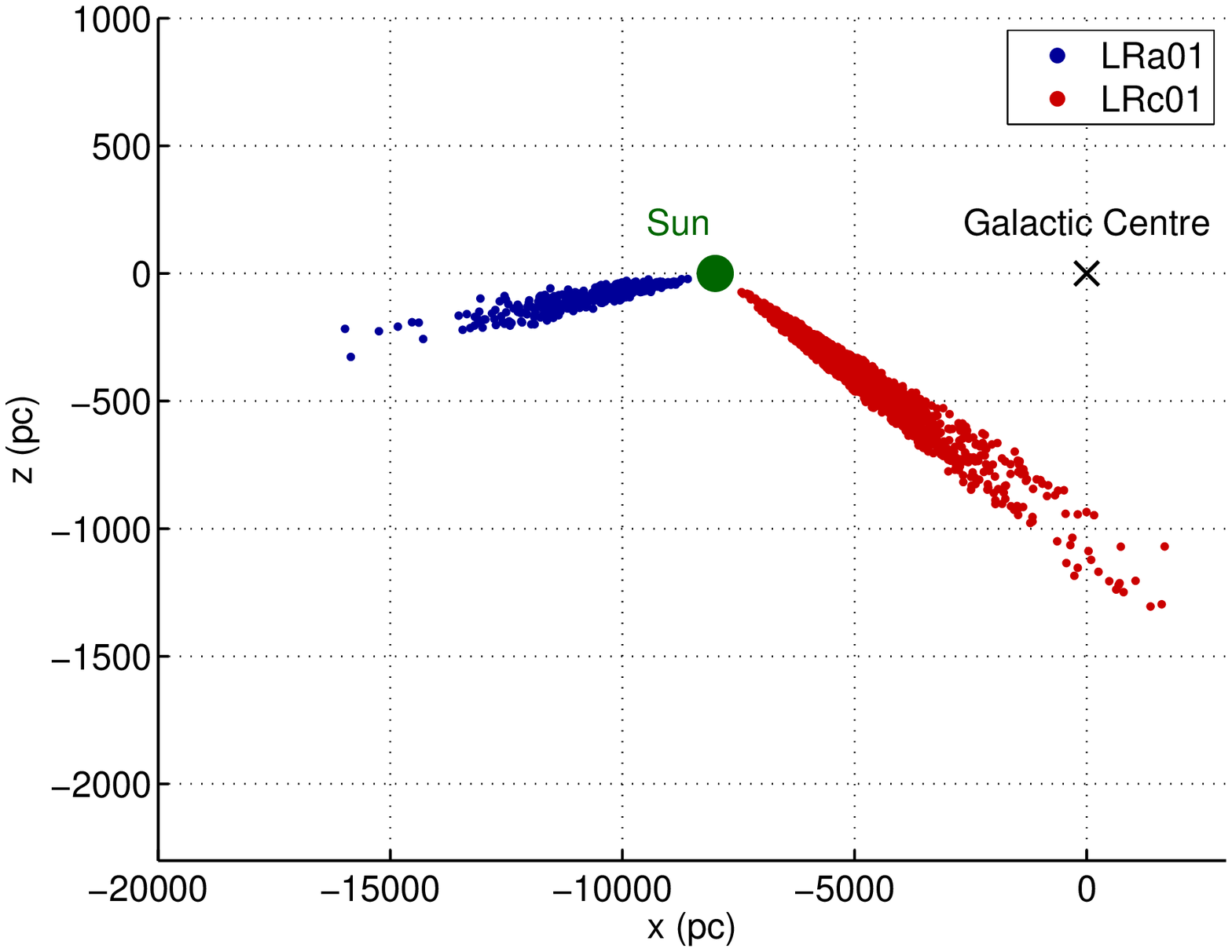}\hspace{1.5cm}
      \includegraphics[width=.37\hsize]{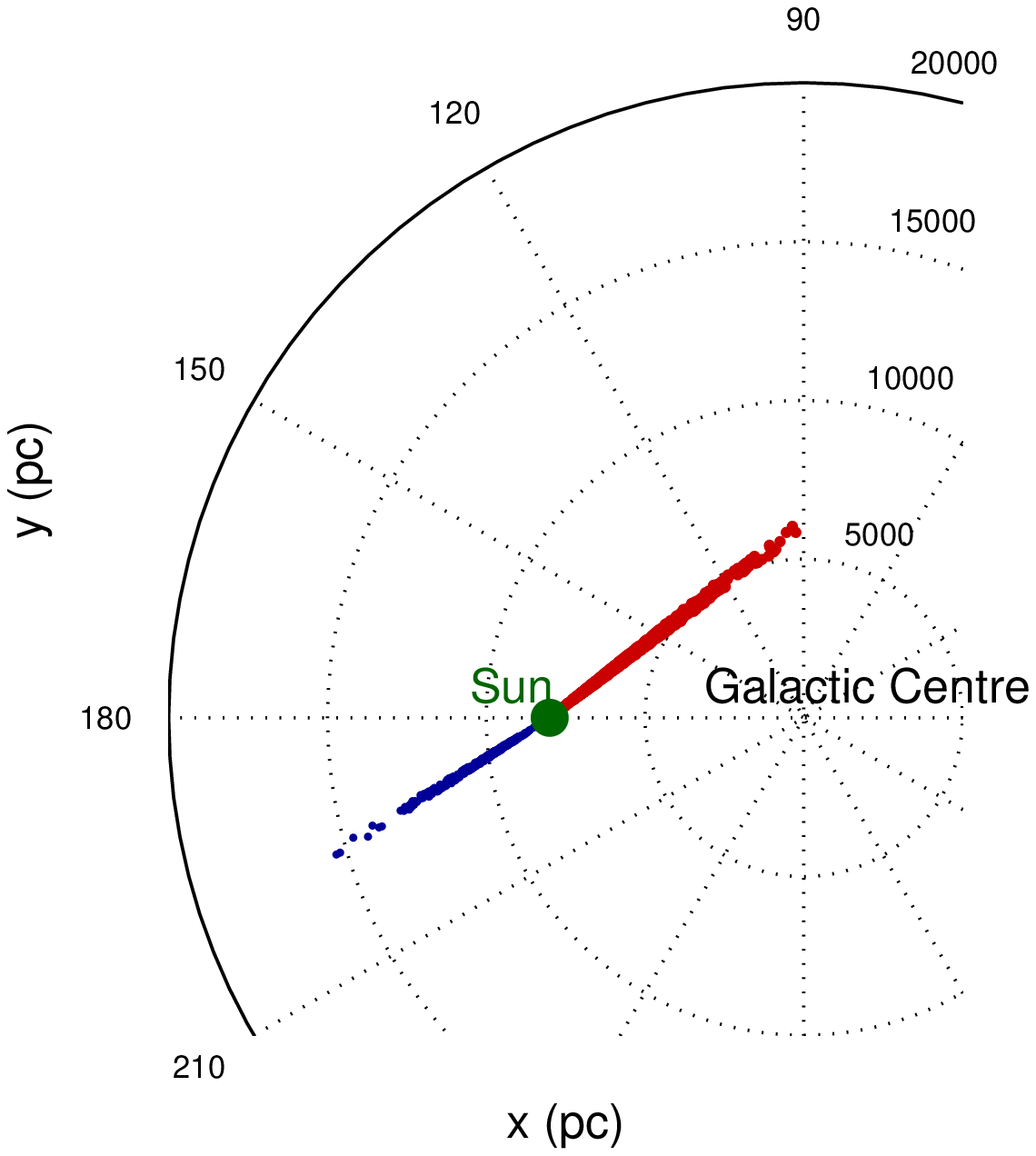}

      \caption{Positions of red giants in the two observed fields (CoRoT designations LRc01 and LRa01) projected on the Galactic plane (right panel) and on a plane perpendicular to the disc (left panel).}
         \label{fig:dist}
\end{figure*}

The positions of the giants in the LRc01 and LRa01 are illustrated in Fig. \ref{fig:dist}.

\section{Differential population study}
Before comparing populations of giants observed in the two fields, we checked whether different biases were introduced in the target selection in LRa01 and LRc01.  
We retrieved from EXODAT \citep{Deleuil2009} photometric information on all the stars in the field, as well as the targets observed. Targets within each field of view were selected largely on the basis of colour-magnitude criteria. 
Within the observed targets, solar-like oscillations were searched for in stars belonging to a limited colour-magnitude domain: $0.6 < J-K_s <1 $ and $K_s < 12$ \citep[see][]{Mosser2010}. Restricting to this domain, we find no significant difference in the target selection bias applied to LRc01 compared to LRa01 (see also \citealt{Miglio2013}). 

\begin{figure}
\vspace*{-0.2cm}

\centering
      \includegraphics[width=.65\hsize]{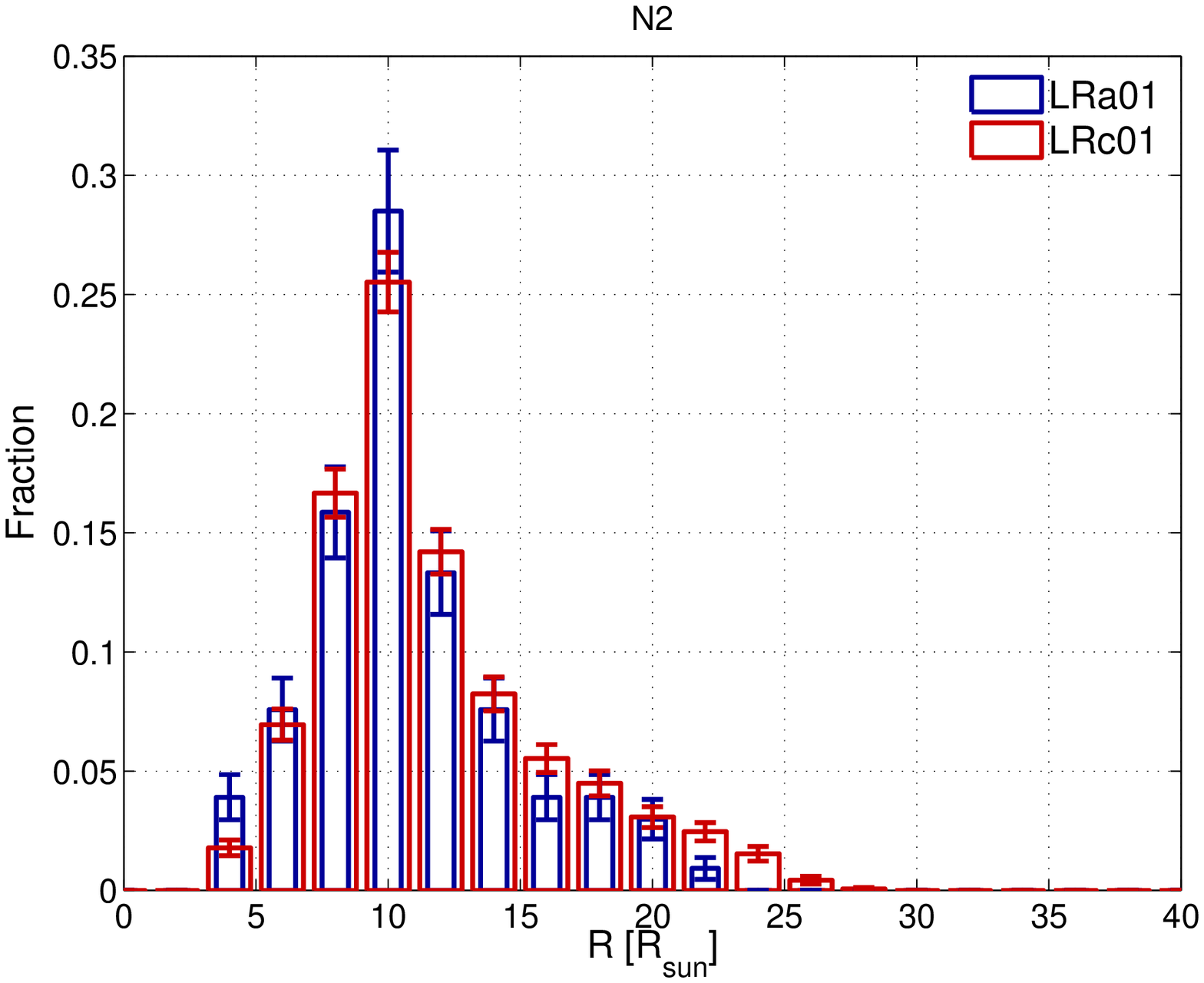}
      \includegraphics[width=.65\hsize]{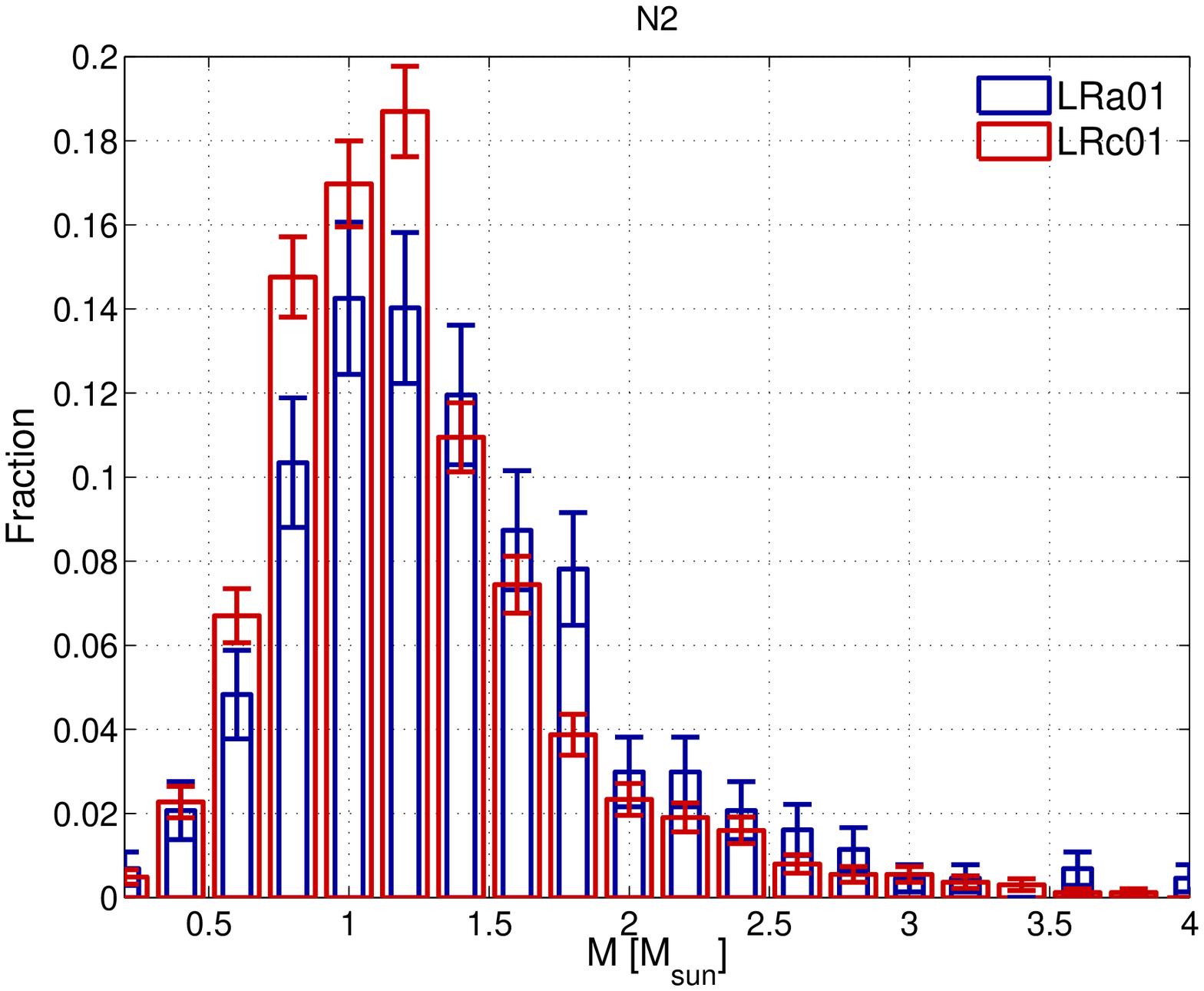}

      \caption{Radius and mass distributions of stars in LRa01 (blue) and LRc01 (red) obtained using the method described in Sec. \ref{sec:dir}}.         \label{fig:N3N2direct}
\end{figure}

 We then compare the distributions of radius and mass of stars in LRa01 and LRc01.
Since scaling relations were not tested at high luminosities we excluded from the sample the few stars with $L>200$ L$_\odot$, and stars with no 2MASS photometry available. The fraction of targets excluded represents $< 3\%$ of the whole sample.
The distributions of $R$ and $M$ are reported in Fig. \ref{fig:N3N2direct}.

We applied the Kolmogorov-Smirnov (K-S) test to the distributions of mass and radius to quantify differences between the two populations. 
When comparing radii, the null hypothesis (LRa01 and LRc01 samples  are drawn from the same parent distribution) cannot be rejected, or the difference between the two populations is marginally significant at best. On the other hand, we find the difference between the mass distribution of the two populations to be highly significant (K-S probability higher than 99.9 \%).  

To test the impact on the K-S test of statistical uncertainties on mass and radius, we perturbed the observed radii and masses adding random offsets  drawn from a Gaussian distribution having a standard deviation equal to the estimated uncertainty on the mass / radius.
We generated 1000 realisations and performed a K-S test on each realisation. The distribution of results from the 1000 K-S tests confirms that while comparing radius  the null hypothesis cannot be rejected, the difference in the mass distribution is significant (in 95\% of the realisations the K-S probabilities are higher than 99\%) .
 
\subsection{Comparison with synthetic stellar populations}
For a meaningful comparison between observed and synthetic populations, we applied to the synthetic population the same selection criteria based on colour and magnitude adopted in the target selection. Moreover, as discussed in detail in \citet{Mosser2010}, since no significant bias in the distribution of the targets is present in the  \numax\, range between 6 $\mu$Hz and 80 $\mu$Hz,  we only considered observed and simulated stars with \numax\ in this frequency range. 
We considered synthetic populations computed with two codes: the Besan\c{c}on Model of the Milky Way \citep{Robin03}, and with TRILEGAL \citep{Girardi05}.

\begin{figure*}
\centering
\vspace*{-0.5cm}
      \includegraphics[width=.9\hsize]{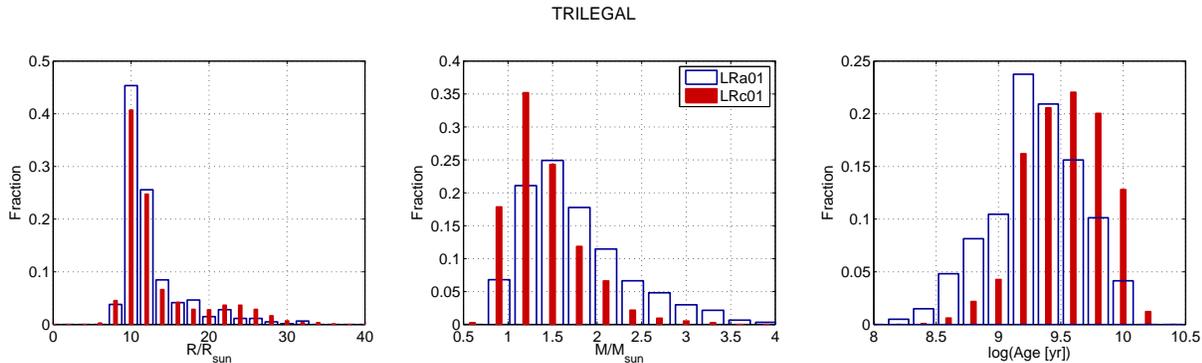}

      \caption{TRILEGAL simulations. Distributions of radius ({\it left panel}), mass ({\it middle panel}), age  ({\it right panel}) of giants in synthetic populations computed with TRILEGAL for stars in the field LRa01 (blue) and LRc01 (red). Stars in the synthetic population were selected in magnitude ($K_s < 12$), colour ($0.6 < J-K_s <1 $), and \numax\, (6 $\mu$Hz $<$ \numax\, $<$ 80 $\mu$Hz) to account for target selection effects.}
         \label{fig:trile}
\end{figure*}

Simulations with both TRILEGAL (see Fig. \ref{fig:trile})  and  the Besan\c{c}on Model show that, although a similar distribution of radius is expected in LRa01 and LRc01, the age (hence mass) distribution of the two populations of giants is  expected to differ. Stars in LRc01 are expected to be older, on average, than those in LRa01.  Additional tests presented in \citet{Miglio2013} show that, in the simulations, the different mass distribution expected in the two fields is indeed related to the assumed age-dependent vertical scale height of the thin disc.
Our aim here is not to find the synthetic population that best matches the data, but to show that the difference we see in the observed distribution is in qualitative agreement with the simulations which include an increase of the disc scale height with age.

The estimated masses of our sample provide important constraints on the stellar ages. Once a star has evolved to the red-giant phase, its age is determined to good approximation by the time spent in the core-hydrogen burning phase, and this is predominantly a function of mass. The CoRoT giants cover a mass range from $\sim0.9$ to $\sim3$ M$_\odot$, which in turn implies an age range spanning $\sim 0.3$ to $\sim 12$ Gyr, i.e., the entire Galactic history. 
To estimate the age of giants in the two observed populations we use  PARAM, a Bayesian stellar parameter estimation method described in detail in  \citet{DaSilva2006}. For this work we adapted PARAM\footnote{This extended version of PARAM will  be made available, via an interactive web form, at the URL {\tt http://stev.oapd.inaf.it/param}.} to include as additional observational constraints \numax, \Dnu\ {and, when available, the evolutionary status of the star (core-Helium or hydrogen-shell burning phase), as determined by the period spacing of gravity-dominated modes  \citep[see][]{Mosser2011}. Having knowledge of the evolutionary status is particularly useful in the age determination of stars with an estimated radius typical of RC giants. Since the mass of core-He burning stars is likely to be affected by mass loss, the age-mass relation in red-clump stars may be different from that of RGB stars (see e.g. Fig. 1 in \citealt{Miglio2012c}). }

We applied this parameter estimation method to the sample of giants observed in LRc01 and LRa01 and obtained distributions of age, mass and radius.
The estimated uncertainty on the age is of the order of 30-40\% given that no information on the metallicity is available, hence a broad prior on $\rm[Fe/H]$ is assumed . 
The results obtained with a model-based approach support the interpretation that the differences between the observed population in LRa01 and LRc01 are due to a different mass, hence age, distribution.  


\section{Conclusions}

In this paper we have shown that solar-like oscillating giants are key tracers of stellar populations in the Milky Way. When combined with photometric constraints, the pulsation spectra of solar-like oscillating giant stars not only reveal their radii, and hence distances, but also provide well-constrained estimates of their masses, which are reliable proxies for the ages of the stars.

{We considered} red giants observed by the CoRoT space telescope in two regions located at different positions on the sky, with CoRoT designations LRc01 ($l=37^\circ$, $b=-7^\circ$), and LRa01 ($l=217^\circ$, $b=-2^\circ$).
We found a significant difference in the mass distributions of the populations of the stars in the two fields, with stars in LRc01 having a lower average mass than those in LRa01. Differences in radius are in contrast marginal at best. To interpret these findings we have compared the observed distributions with the distributions given by synthetic population simulations  of red giants  for the same fields of the Galaxy covered by the observations. 
The synthetic populations show a difference in the mass distributions that agrees qualitatively with that found in the observed populations.

On the basis of these comparisons we interpret the differences in the mass distributions as being due mainly to the different average heights of the observed fields below the galactic plane. 
Since it is believed that dynamical processes in the disc increase the velocity dispersion of stars with time, it follows that older stellar populations reach greater heights above and below the plane. What we observe with CoRoT, i.e. that the field higher below the plane has a larger fraction of old (i.e low mass) stars, thus seems to support this theoretical expectation. Other data may hint at a similar dependence, as for instance an increase of the velocity dispersion of stars with increasing age \citep{Holmberg2009}, or the correlation between scale height and [O/Fe]  \citep{Bovy2012}, but they are only available for stars in the solar vicinity.
Future analyses of pulsating red giants sampling the Milky Way at different radii and heights from the plane, when complemented with their chemical abundances, will set even tighter constraints and enable to fully disentangle both the radial and vertical structure of the Galactic disc. 

The findings presented here {provide an}  example of the detailed picture of Galactic populations inferred from pulsating red giants. We can map regions that are more than a factor ten further away in distance compared to what has previously been possible, and we can explore (with a homogeneous sample) a wide age interval sampling look-back times as long as the age of the Galaxy. 
This method, when applied to the various regions explored by CoRoT and \kepler, and complemented by chemo-dynamical constraints from spectroscopic analyses, will provide the gold standard for current and future surveys of the Milky Way.

\section*{Acknowledgements}
TM acknowledges financial support from Belspo for contract PRODEX GAIA-DPAC. JM and MV acknowledge financial support from Belspo for contract PRODEX COROT.
\bibliographystyle{mn2e_new}
\vspace*{-0.6cm}
\small
\bibliography{andrea_m}
\label{lastpage}
\end{document}